# Spectroscopic and Photometric Study of the Asymptotic Giant Branch Star T Cephei

**David Boyd**
*West Challow Observatory, OX12 9TX, UK; davidboyd@orion.me.uk*

**Abstract**   Spectroscopy and photometry of the Asymptotic Giant Branch star T Cephei were recorded concurrently on 36 nights during its 387-day pulsation cycle in 2022. Photometry was used to calibrate all spectra in absolute flux. We report on the variation of B and V magnitudes, B–V color index, spectral type, effective temperature, and Balmer emission line flux during one complete pulsation cycle.

## 1. Introduction

T Cep is an oxygen-rich Mira star discovered by Ceraski in 1880 (Schmidt 1881). Miras are red giant stars with spectral type M on the Asymptotic Giant Branch (AGB) of the Hertzsprung-Russell diagram (Percy 2007). They are in the final stages of their lives prior to becoming planetary nebulae and eventually white dwarfs. Pulsation in their atmospheres with a typical period of around a year drives mass loss through a slow wind and forms a tenuous outer atmosphere. The temperature in the atmosphere is low enough for molecules such as TiO to form. These molecules absorb light from the stellar continuum in the visual part of the spectrum, causing TiO molecular absorption bands which are a prominent feature in the spectrum of oxygen-rich Miras. During each pulsation cycle the star brightens as more light is emitted in the visual part of the spectrum and its effective temperature rises. This causes some of the TiO molecules to dissociate, reducing the strength of the molecular absorption bands and making its apparent spectral type earlier. As the star fades, it becomes cooler, redder, molecules reform, and its spectral type becomes later. Phase zero of the pulsation cycle in Miras is normally taken as the time of maximum brightness. In some oxygen-rich Miras, emission lines of the hydrogen Balmer series appear around this time then gradually disappear. A comprehensive review of our knowledge about Mira stars is given in Willson and Marengo (2012).

A short introduction to T Cep is given in the "Star of the Year" article in the 2023 *BAA Handbook* (BAA 2023c). An article about T Cep in the December 2021 issue of the BAA Variable Star Circular (Heywood 2021) highlighted its complex behavior during the rise to maximum in recent pulsation cycles (see Figure 1) and invited spectroscopic observations of T Cep as there appeared to be relatively little spectroscopic data available on the star. In response to this invitation, spectra of T Cep were recorded on 36 nights between December 2021 and January 2023, covering the complete 2022 pulsation cycle. B and V magnitudes were measured concurrently with the spectra.

## 2. Determining the current pulsation cycle epoch and phase

The pulsation cycle of T Cep has been shown to vary in period and amplitude by Isles and Saw (1989). In order to find the current pulsation period, we downloaded V-band magnitude data for the 2019 to 2022 pulsation cycles from the AAVSO International Database (AID; Kloppenborg 2022) as shown in Figure 1. These include data reported by the author for the 2022 cycle and analyzed here. To establish parameters of the current pulsation cycle, we carried out period analyses of these data in groups of cycles using the ANOVA method in Peranso (Paunzen and Vanmunster 2016). Table 1 shows how the mean pulsation period varies depending on how many cycles are included in the analysis. The most recent period of 387 days has been adopted as the current pulsation period for the purpose of calculating the pulsation phase of our 2022 observations. Given the irregular profile of the peak in recent cycles, we used the more clearly defined minima as the basis for defining the epoch and phase of the 2022 cycle. A quadratic polynomial fitted to the lower part of the minimum at the start of this cycle gave the epoch of phase –0.5 as JD = 2459553.2(3).

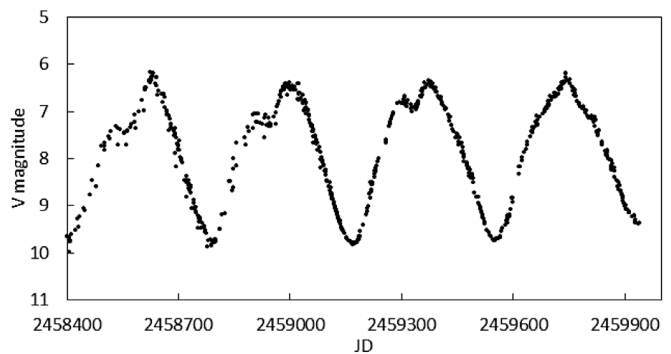

Figure 1. V-magnitude light curve for the 2019 to 2022 pulsation cycles from the AAVSO International Database (cleaned to remove obvious outliers).

Table 1. Fitted pulsation periods for recent pulsation cycles.

| Cycles included in analysis | Fitted pulsation period (d) |
|---|---|
| 2019–2022 | 382 ± 2 |
| 2020–2022 | 385 ± 3 |
| 2021–2022 | 387 ± 4 |

## 3. Equipment and data reduction

Spectra of T Cep were obtained with a 0.28-m Schmidt-Cassegrain Telescope (SCT) operating at f/5 equipped with an auto-guided Shelyak LISA slit spectrograph and a SXVR-H694

CCD camera. The slit width was 23μ, giving a mean spectral resolving power of ~1000. Spectra were processed with the ISIS spectral analysis software (Buil 2021). Spectroscopic images were bias, dark, and flat corrected, geometrically corrected, sky background subtracted, spectrum extracted, and finally wavelength calibrated using the integrated ArNe calibration source. Each T Cep spectrum was then corrected for instrumental and atmospheric losses by recording the spectrum of a nearby reference star with a known spectral profile from the MILES library of stellar spectra (Falcón-Barroso *et al.* 2011). Each reference star was chosen as close as possible in airmass to T Cep and its spectrum obtained immediately prior to the T Cep spectrum. Typically, ten five-minute guided integrations were recorded for each spectrum of T Cep, giving signal-to-noise ratios ranging from 110 at maximum brightness to 70 at minimum. Spectra were calibrated in absolute flux in FLAM units as erg/cm$^2$/s/Å using the V magnitudes measured concurrently with the spectra as described in Boyd (2020). All spectra were submitted to, and are available from, the BAA Spectroscopy Database (BAA 2023b).

The distance to T Cep according to Gaia DR2 is 176 –12 +14 parsecs (Gaia Collaboration *et al.* 2018). According to Schlafly and Finkbeiner (2011), the total galactic extinction in the direction towards T Cep is E(B–V) = 0.057. As T Cep lies relatively nearby at galactic latitude +13°, the interstellar extinction it experiences is likely to be considerably smaller than this. Therefore no correction for interstellar reddening is applied to spectra of T Cep.

Each night, while spectra were being recorded, photometry of T Cep was obtained with a 0.35-m SCT operating at f/5 equipped with Astrodon Johnson-Cousins photometric filters and a Starlight Xpress SXVR-H9 CCD camera. All photometric observations were made through alternating B and V filters with typically 10 to 15 images recorded in each filter. Photometric images were bias, dark, and flat corrected and instrumental magnitudes obtained by aperture photometry using the software AIP4WIN (Berry and Burnell 2005). An ensemble of five nearby comparison stars was used whose B and V magnitudes were obtained from the AAVSO Photometric All-Sky Survey (APASS; Henden *et al.* 2021). Instrumental B and V magnitudes were transformed to the Johnson UBV photometric standard using the measured B–V color index and atmospheric airmass with the algorithm published in Boyd (2011). As the star approached peak brightness, exposures had to be shortened to ensure the CCD camera continued to operate in the linear region. These shorter exposures resulted in larger uncertainties on the individual measurements because of scintillation. Magnitude measurements for each night were averaged and nightly mean B and V magnitudes were converted to absolute B- and V-band flux using photometric zero points derived from CALSPEC spectrophotometric standard stars (Bohlin *et al.* 2014; STScI 2021). Times are recorded as Julian Date (JD). All measured magnitudes were submitted to, and are available from, the BAA Photometry Database (BAA 2023a) and nightly means are available in the AAVSO International Database.

## 4. Photometric observations

Julian Date, pulsation phase, nightly mean B and V magnitudes, and B–V color index for T Cep during the 2022 pulsation cycle are listed in Table 2. Uncertainties in nightly means in B and V range from 0.013 around minimum to 0.043 around maximum. These uncertainties are propagated into the uncertainties in the B–V color index and absolute flux. Nightly mean B and V magnitudes, B–V color indices, and B- and V-band absolute fluxes are plotted vs phase in Figure 2. Uncertainties in B and V magnitudes in Figure 2 are within the plotted symbols.

The pronounced dip in the light curve as the star approaches maximum seen in the previous three years is absent in 2022 and replaced by a flattening of the slope in magnitude and steepening of the slope in flux. While the B and V magnitudes have a broad peak at phase zero, flux peaks more sharply. Figure 3 shows the correlation between B–V color index and V magnitude through the pulsation cycle. The B–V color index reaches its bluest before the magnitude peaks, approximately at the phase where the magnitude dipped in previous years, then reddens as

Table 2. Julian Date, pulsation phase, nightly mean B and V magnitude and B–V color index for T Cep during the 2022 pulsation cycle.

| Julian Date | Phase | B (mag) | V (mag) | B–V (mag) |
| --- | --- | --- | --- | --- |
| 2459570 | –0.46 | 11.49 | 9.49 | 2.00 |
| 2459584 | –0.42 | 11.19 | 9.22 | 1.96 |
| 2459597 | –0.39 | 10.77 | 8.85 | 1.92 |
| 2459617 | –0.33 | 9.95 | 8.17 | 1.78 |
| 2459637 | –0.28 | 9.35 | 7.66 | 1.69 |
| 2459653 | –0.24 | 8.98 | 7.33 | 1.64 |
| 2459661 | –0.22 | 8.84 | 7.23 | 1.61 |
| 2459665 | –0.21 | 8.80 | 7.19 | 1.61 |
| 2459672 | –0.19 | 8.71 | 7.12 | 1.59 |
| 2459685 | –0.16 | 8.48 | 6.96 | 1.51 |
| 2459694 | –0.14 | 8.34 | 6.80 | 1.54 |
| 2459704 | –0.11 | 8.22 | 6.70 | 1.52 |
| 2459711 | –0.09 | 8.17 | 6.61 | 1.56 |
| 2459721 | –0.07 | 8.12 | 6.52 | 1.60 |
| 2459742 | –0.01 | 7.92 | 6.27 | 1.65 |
| 2459753 | 0.02 | 8.08 | 6.33 | 1.75 |
| 2459762 | 0.04 | 8.20 | 6.50 | 1.70 |
| 2459770 | 0.06 | 8.37 | 6.68 | 1.70 |
| 2459779 | 0.08 | 8.61 | 6.85 | 1.76 |
| 2459797 | 0.13 | 8.64 | 6.99 | 1.65 |
| 2459804 | 0.15 | 8.84 | 7.14 | 1.70 |
| 2459811 | 0.17 | 8.85 | 7.17 | 1.68 |
| 2459821 | 0.19 | 8.97 | 7.31 | 1.66 |
| 2459840 | 0.24 | 9.31 | 7.64 | 1.67 |
| 2459852 | 0.27 | 9.60 | 7.88 | 1.72 |
| 2459859 | 0.29 | 9.74 | 8.02 | 1.72 |
| 2459870 | 0.32 | 10.00 | 8.23 | 1.77 |
| 2459881 | 0.35 | 10.28 | 8.48 | 1.80 |
| 2459896 | 0.38 | 10.68 | 8.81 | 1.87 |
| 2459904 | 0.40 | 10.83 | 8.99 | 1.84 |
| 2459912 | 0.43 | 11.00 | 9.12 | 1.88 |
| 2459921 | 0.45 | 11.14 | 9.24 | 1.90 |
| 2459928 | 0.47 | 11.25 | 9.30 | 1.95 |
| 2459940 | 0.50 | 11.31 | 9.33 | 1.98 |
| 2459954 | –0.47 | 11.17 | 9.19 | 1.98 |
| 2459961 | –0.45 | 10.98 | 9.03 | 1.95 |
| 2459962 | –0.45 | 10.95 | 9.00 | 1.96 |

*Note: Uncertainties in nightly means in B and V range from 0.013 around minimum to 0.043 around maximum.*

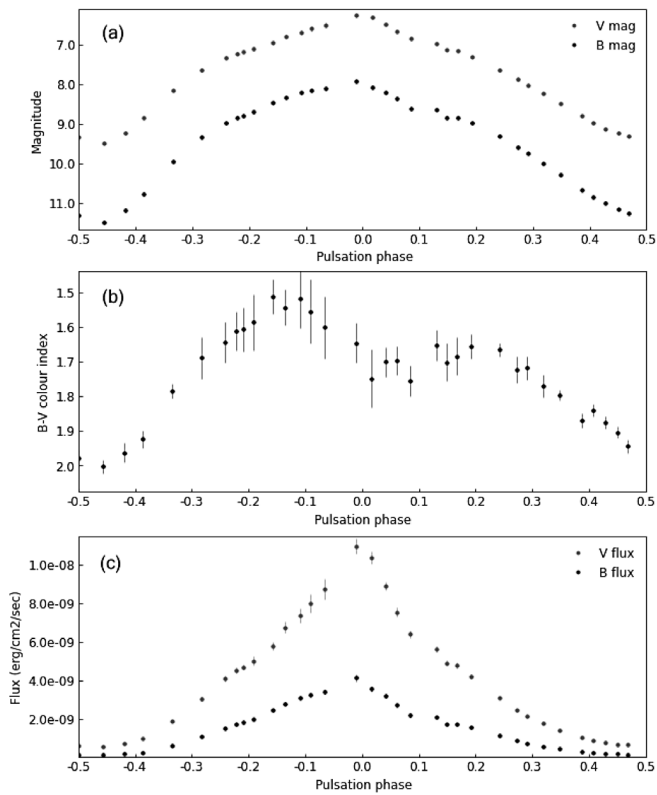

Figure 2. Variation of nightly means of (a) B and V magnitude, (b) B–V color index, and (c) V and B absolute flux with phase during the 2022 pulsation cycle of T Cep. Uncertainties in B and V magnitude are within the plotted symbols.

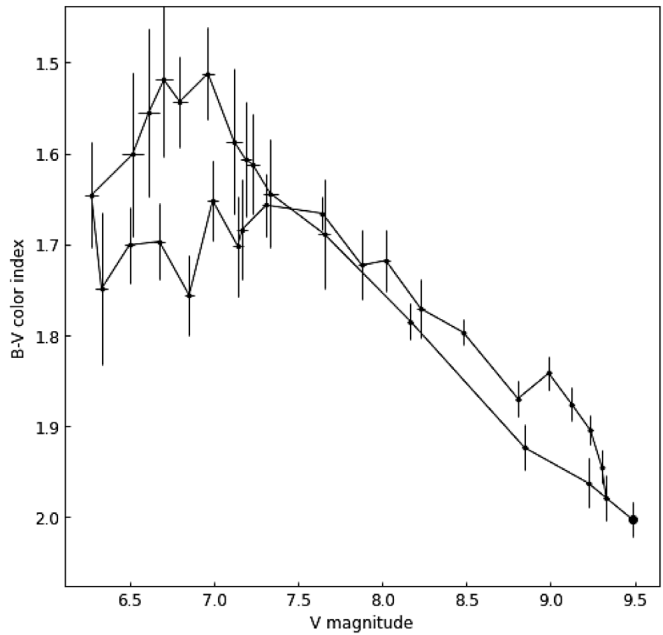

Figure 3. Correlation between B–V color index and V magnitude through the 2022 pulsation cycle of T Cep. The dot marks the beginning of the cycle.

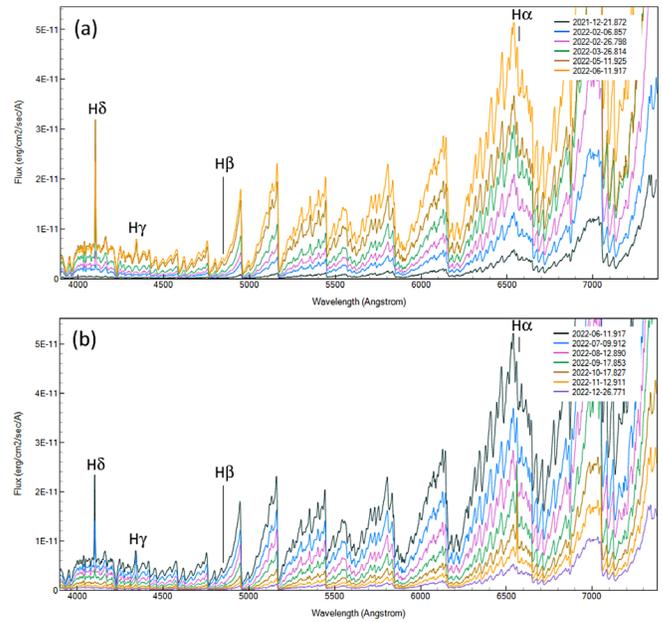

Figure 4. Selected absolute flux calibrated spectra of T Cep (a) on the rise to maximum and (b) on the decline during the 2022 pulsation cycle. The locations of hydrogen Balmer emission lines are marked. The Hδ line is most prominent followed by Hγ, while Hβ and Hα are indistinguishable on this scale.

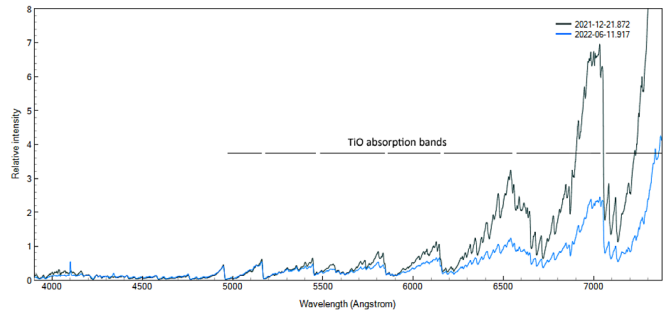

Figure 5. Relative flux spectra of T Cep close to minimum brightness at phase 0.54 (black) and maximum brightness at phase –0.01 (blue), with fluxes mutually normalized in the range 4500 to 5500 Å. The TiO absorption bands indicated are proportionally shallower with increasing wavelength at maximum brightness.

the magnitude peaks before becoming slightly bluer again and finally reddening as the star fades and the cycle ends.

## 5. Spectroscopic observations

Figure 4 shows two composite plots of selected spectra of T Cep calibrated in absolute flux during the rise and the fall of the 2022 pulsation cycle. The location of hydrogen Balmer emission lines are marked. The M giant spectra are punctuated by TiO absorption bands where molecules in the star's atmosphere absorb some of the light emitted from deeper in the star. These plots are deceptive as they appear to show the TiO bands becoming deeper as the star brightens. To get a truer picture of the relative depth of the bands, Figure 5 shows relative flux spectra close to minimum brightness at phase 0.54 (black) and maximum brightness at phase –0.01 (blue) with fluxes mutually normalized in the range 4500 to 5500 Å. This shows that the TiO absorption bands indicated are proportionally shallower with increasing wavelength at maximum brightness.

Table 3. Julian Date, spectral sub-type, effective temperature ($T_{eff}$) and Balmer line absolute flux for T Cep spectra recorded during the 2022 pulsation cycle.

| Julian Date | Spectral Sub-Type | $T_{eff}$ (K) | Hα Line Flux (ergs/cm$_2$/sec) | Hβ Line Flux (ergs/cm$_2$/sec) | Hγ Line Flux (ergs/cm$_2$/sec) | Hδ Line Flux (ergs/cm$_2$/sec) |
|---|---|---|---|---|---|---|
| 2459570 | M9.4 | 2630 | 0.00E+00 | 0.00E+00 | 0.00E+00 | 1.67E–13 |
| 2459584 | M9.0 | 2718 | 6.41E–14 | 0.00E+00 | 0.00E+00 | 2.86E–13 |
| 2459597 | M8.4 | 2845 | 0.00E+00 | 0.00E+00 | 0.00E+00 | 9.72E–13 |
| 2459617 | M8.1 | 2906 | 0.00E+00 | 0.00E+00 | 0.00E+00 | 4.23E–12 |
| 2459637 | M8.0 | 2926 | 5.59E–13 | 0.00E+00 | 3.44E–14 | 1.39E–11 |
| 2459653 | M8.0 | 2926 | 2.50E–12 | 0.00E+00 | 6.98E–13 | 2.56E–11 |
| 2459661 | M8.0 | 2926 | 0.00E+00 | 0.00E+00 | 1.76E–12 | 5.11E–11 |
| 2459665 | M8.0 | 2926 | 0.00E+00 | 0.00E+00 | 2.40E–12 | 5.52E–11 |
| 2459672 | M7.9 | 2946 | 0.00E+00 | 0.00E+00 | 3.28E–12 | 6.24E–11 |
| 2459685 | M7.8 | 2965 | 0.00E+00 | 6.86E–14 | 5.39E–12 | 7.15E–11 |
| 2459694 | M7.8 | 2965 | 0.00E+00 | 1.65E–13 | 7.77E–12 | 7.03E–11 |
| 2459704 | M7.5 | 3022 | 0.00E+00 | 6.58E–15 | 1.39E–11 | 1.05E–10 |
| 2459711 | M7.5 | 3022 | 8.49E–13 | 2.40E–13 | 1.68E–11 | 1.23E–10 |
| 2459721 | M7.6 | 3003 | 1.56E–12 | 5.53E–13 | 1.66E–11 | 1.07E–10 |
| 2459742 | M7.5 | 3022 | 9.90E–12 | 1.69E–12 | 2.50E–11 | 8.12E–11 |
| 2459753 | M7.5 | 3022 | 1.85E–11 | 2.97E–12 | 2.82E–11 | 7.46E–11 |
| 2459762 | M7.5 | 3022 | 2.31E–11 | 3.40E–12 | 2.78E–11 | 6.24E–11 |
| 2459770 | M7.5 | 3022 | 2.49E–11 | 2.86E–12 | 2.36E–11 | 4.43E–11 |
| 2459779 | M7.6 | 3003 | 2.22E–11 | 2.49E–12 | 1.52E–11 | 2.31E–11 |
| 2459797 | M7.7 | 2984 | 1.80E–11 | 1.77E–12 | 6.81E–12 | 5.37E–12 |
| 2459804 | M7.7 | 2984 | 1.56E–11 | 1.43E–12 | 4.92E–12 | 2.42E–12 |
| 2459811 | M7.9 | 2946 | 2.16E–11 | 1.11E–12 | 4.64E–12 | 2.44E–12 |
| 2459821 | M7.8 | 2965 | 2.16E–11 | 1.56E–12 | 6.52E–12 | 4.91E–12 |
| 2459840 | M7.9 | 2946 | 4.70E–11 | 3.53E–12 | 8.84E–12 | 5.93E–12 |
| 2459852 | M7.9 | 2946 | 5.30E–11 | 3.83E–12 | 6.45E–12 | 2.41E–12 |
| 2459859 | M8.0 | 2926 | 5.13E–11 | 3.36E–12 | 4.42E–12 | 1.06E–12 |
| 2459870 | M8.0 | 2926 | 4.81E–11 | 3.15E–12 | 2.67E–12 | 5.38E–13 |
| 2459881 | M8.2 | 2886 | 4.41E–11 | 2.85E–12 | 1.61E–12 | 7.85E–14 |
| 2459896 | M8.3 | 2866 | 2.93E–11 | 1.99E–12 | 6.55E–13 | 0.00E+00 |
| 2459904 | M8.7 | 2782 | 2.22E–11 | 1.48E–12 | 4.73E–13 | 1.24E–13 |
| 2459912 | M8.6 | 2804 | 1.58E–11 | 1.08E–12 | 3.17E–13 | 0.00E+00 |
| 2459921 | M8.2 | 2886 | 9.53E–12 | 7.58E–13 | 8.37E–14 | 0.00E+00 |
| 2459928 | M8.4 | 2845 | 7.16E–12 | 4.73E–13 | 2.61E–14 | 0.00E+00 |
| 2459940 | M9.0 | 2718 | 3.79E–12 | 2.19E–13 | 6.25E–15 | 0.00E+00 |
| 2459954 | M8.9 | 2740 | 1.72E–12 | 1.01E–13 | 0.00E+00 | 0.00E+00 |
| 2459962 | M9.0 | 2718 | 1.51E–12 | 6.87E–14 | 0.00E+00 | 8.72E–14 |

*Note: Line flux too small to measure is shown as zero. Estimated uncertainty in spectral sub-type is ± 0.2, in $T_{eff}$ is ± 40 K and in line flux is ± 15%.*

### 6. Measuring absolute flux in the Balmer emission lines

The presence of hydrogen emission lines in the spectrum of o Ceti was first mentioned in 1887 by Pickering at Harvard College Observatory (Pickering 1887). Observation of Hδ and Hγ emission in the spectra of Mira stars around maximum brightness was subsequently reported by, among others, Merrill (1921), Frost and Lowater (1923), and Joy (1926), although their cause was not understood. At that time observation of the longer wavelength Balmer lines was hampered by the low red sensitivity of photographic plates. Observation of Hα emission lines using small telescopes with objective prisms and photographic plates chemically sensitized to red light was encouraged by Merrill (1920).

Current understanding is that the hydrogen Balmer emission lines in oxygen-rich Mira stars are caused by shock waves generated deep in the star's atmosphere below the level of molecular absorption as the outward pressure of radiation is countered by the inward pressure of gravity. These shock waves propagate radially outwards, ionizing hydrogen in the atmosphere and driving mass loss. Recombination then generates emission lines that rise above the M-giant continuum as the star passes through its pulsation cycle (Willson 1976; Gillet *et al.* 1983).

To measure the absolute flux in an emission line, we have to subtract the contribution to the flux at that wavelength from the M-giant continuum. Here we describe the process adopted for the Hα line. To establish the M-giant continuum under the Hα Balmer line, we selected those absolute flux spectra which clearly showed no additional emission at that wavelength. We then averaged these spectra to construct a reference continuum spectrum under the Hα line. For each absolute flux spectrum with visible Hα emission, we scaled this Hα reference spectrum so it aligned in flux with the profile of that absolute flux spectrum in regions on either side of the Hα line. The scaled Hα reference spectrum was then subtracted from the absolute flux spectrum to give the absolute flux profile of only the Hα emission line in that spectrum. This was integrated over the wavelengths of the line to give the absolute flux emitted in the Hα line in that spectrum. This process was repeated for the other Balmer lines and for each spectrum. We checked that this procedure was robust against any small changes in the M-giant continuum during the pulsation cycle. The absolute fluxes found for each Balmer line in each of our spectra are listed in Table 3 and plotted vs pulsation phase in Figure 6. The uncertainties in each line flux measurement in Figure 6 are a combination of uncertainties in determining the absolute flux level of the continuum in the region of the line and in integrating the line profile over the interpolated continuum level. This uncertainty is calculated for each measurement and the average for each Balmer line is between 13% and 17%.

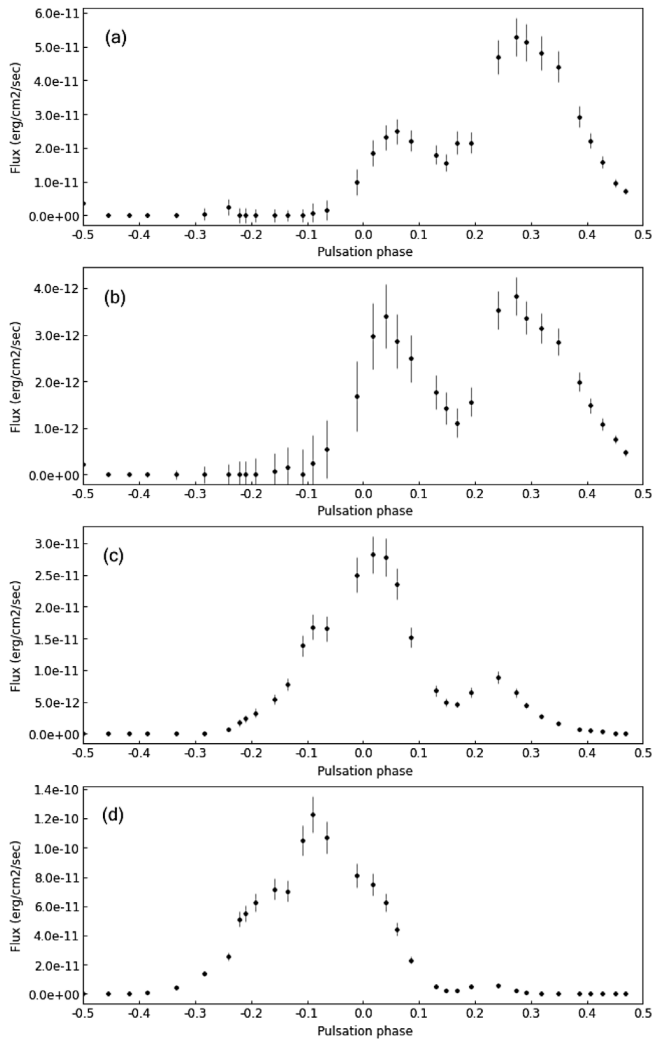

Figure 6. Absolute flux of the (a) Hα, (b) Hβ, (c) Hγ, and (d) Hδ Balmer lines vs pulsation phase. The uncertainties are calculated as described in the text and the average uncertainty for each Balmer line is between 13% and 17%.

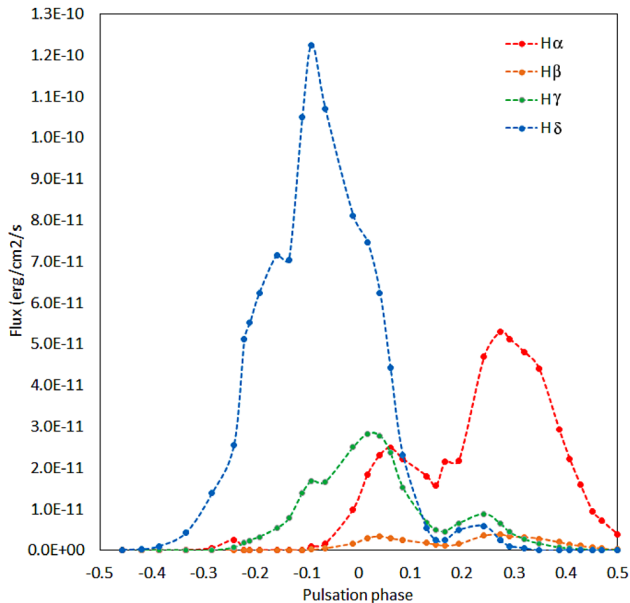

Figure 7. Composite plot showing the variation of absolute flux of the Balmer lines with pulsation phase using a consistent flux scale.

Figure 7 is a composite plot showing the variation of absolute flux of the Balmer lines with pulsation phase using a consistent flux scale. It is noteworthy that the behavior of these lines in T Cep during its 2022 pulsation cycle is very different from that observed in the oxygen-rich Miras RY Cep and SU Cam (Boyd 2021). Whereas in those stars the behavior of all four Balmer lines was broadly similar with a maximum around the phase of maximum V flux, in T Cep flux in Hδ peaks before the maximum of the cycle while Hα peaks much later, and all lines have double peaks. This points to a more complex phase relationship in the atmosphere of T Cep between propagation of the shock wave and generation of each Balmer line.

### 7. Assigning a spectral type and effective temperature to each spectrum

The strength of TiO molecular absorption bands in the spectra of oxygen-rich M-giant stars has been used to estimate spectral type (Wing 1992). Because the strength of the molecular bands changes over the pulsation cycle in Mira stars, the spectral type of the star determined in this way will also change. Assigning a spectral type to an individual spectrum is commonly achieved by comparing it morphologically to a range of standard star spectra in the MK spectral classification system and identifying the closest match (Gray and Corbally 2009). MK standard stars available with the MKCLASS stellar spectral classification system (Gray and Corbally 2014) cover the wavelength range 3800–5600 Å, where atomic absorption lines are concentrated, a legacy of the use of blue-sensitive photographic plates in the early days of the MK standard. In our spectra the flux in this region is relatively low, whereas it is considerably stronger towards the red end of the visual range where the molecular bands are prominent. Given our limited spectral resolution and therefore inability to clearly resolve some of the lines in the blue part of the spectrum used for classification, using the full visual range to classify our spectra offers a more practical and robust way of assigning a spectral type.

As all our T Cep spectra fell within range of the M spectral type, we decided to use the M-giant spectra published in Fluks *et al.* (1994), which are classified on the MK system, to assign an apparent spectral type to each spectrum. The Fluks spectra for spectral types M0 to M10 are defined on the wavelength range 3500–10000 Å at an interval of 1 Å. We normalized the flux of each standard spectrum to a mean flux value of unity in the wavelength interval 5610 Å to 5630 Å, which contains no strong spectral features.

To estimate a spectral type for each of our measured spectra, we obtained the closest match between each of our spectra and M-giant standard spectra in the Fluks Spectral Flux Library. To do this we quadratically interpolated all our spectra onto a 1 Å grid, normalized the flux of each spectrum to a mean flux value of unity in the 5610 Å to 5630 Å wavelength interval, and removed the region of each spectrum around the four Balmer lines. We then computed the total squared flux difference between each of our spectra and each of the standard spectra over the wavelength range 4000 Å to 6600 Å. We identified the spectral type of the standard spectrum which gave the minimum squared flux difference for each of our spectra. We consider this to be the best match integral spectral type. We then fitted a quadratic polynomial to the residuals between the best match integral spectral type and the two adjacent spectral types and used the minimum of this to assign a spectral sub-type to the nearest tenth which best matched each of our spectra. We found the spectral type of T Cep to be M7.5 at phase 0 and M9.4 at phase 0.5. These results are consistent with Keenan *et al.* (1974), who gave the spectral type of T Cep as M6e to M9e.

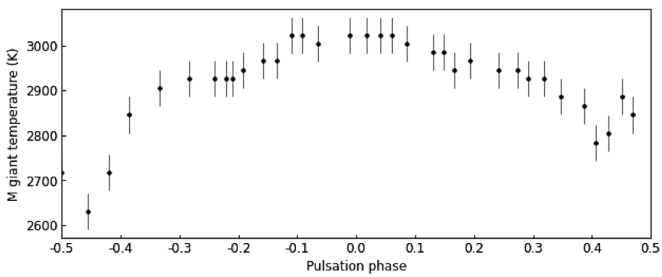

Figure 8. M-giant assigned effective temperatures of T Cep spectra vs pulsation phase.

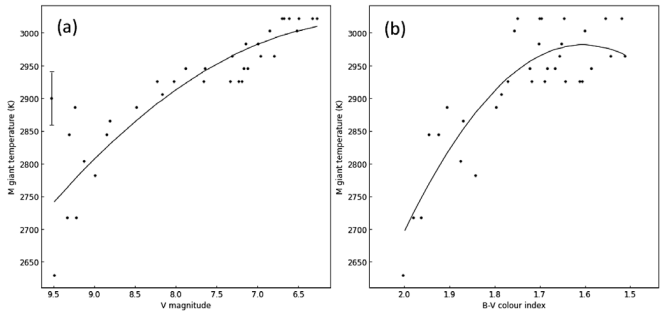

Figure 9. Relationships between M-giant effective temperature of T Cep and (a) V magnitude and (b) B–V color index in the 2022 pulsation cycle. The error bar shows the estimated uncertainty in effective temperature. The lines are quadratic fits to the data but have no physical significance.

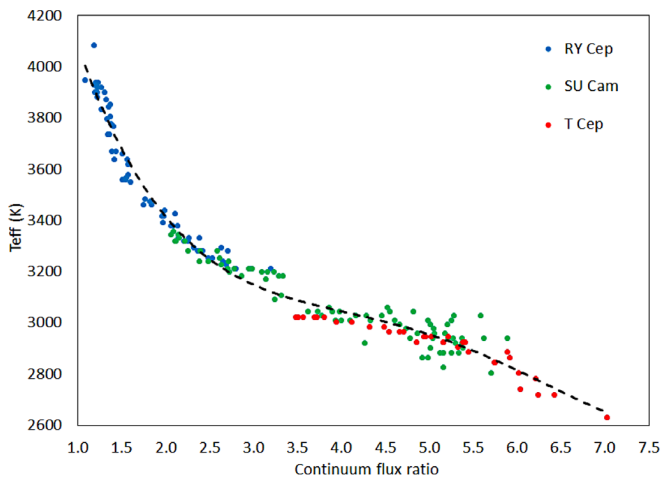

Figure 10. Effective temperature ($T_{eff}$) vs continuum flux ratio for RY Cep (blue), SU Cam (green), and T Cep (red) spectra plus a fitted fourth-order polynomial.

We calculated mean values of effective temperature for M giant spectral types between M6 and M10 from data given in Fluks *et al.* (1994), van Belle *et al.* (1999), and Gray and Corbally (2009). We then fitted a fourth-order polynomial to these mean effective temperatures as a function of spectral type and used this to assign an effective temperature for each of our spectra based on their assigned spectral sub-type. These assigned spectral sub-types and effective temperatures for our T Cep spectra are listed in Table 3 and the effective temperatures plotted vs pulsation phase in Figure 8. We estimate the uncertainty in assigning a spectral sub-type is ± 0.2 and in assigning an effective temperature is ± 40 K.

The relationships between effective temperature of T Cep and (a) V magnitude and (b) B–V color index are shown in Figure 9. This confirms that B–V color index is a relatively poor indicator of effective temperature in Mira stars. This may be partly due to a relatively weak correlation between the B-band region of the spectrum and the longer wavelength region, which has a strong influence on the calculation of effective temperature.

## 8. Estimating effective temperature from continuum flux ratios

Wing (1992) introduced the concept of measuring spectral flux of red variables using three narrow-band interference filters with defined passbands to derive an index of TiO band strength which enabled a spectral type to be assigned. A practical difficulty in using this technique for many amateurs, besides the need to purchase these filters, is that the wavelength regions used in this system extend beyond the range of most spectrographs in use by amateurs.

Inspired by the work of Wing, in Boyd (2021) we devised a simpler approach to estimating spectral type which does not require the use of Wing filters. This involves measuring the mean flux in two narrow wavelength regions, 6130–6140 Å and 6970–6980 Å, and calculating their flux ratio. These regions are included in many amateur spectra and are adjacent to TiO molecular band heads so are likely to be as close as possible to the original unabsorbed photospheric continuum. This flux ratio is in effect measuring a color index based on magnitudes corresponding to mean fluxes in these two spectral regions. Because this involves calculating a flux ratio, it does require the spectrum to be calibrated in relative flux across this spectral range but not necessarily in absolute flux. It also requires any interstellar extinction which would affect the continuum slope to have been corrected. We first applied this approach to the Mira stars RY Cep and SU Cam as reported in Boyd (2021), where we found this calculated continuum flux ratio to be strongly correlated to assigned effective temperature and therefore could be used to estimate effective temperature.

To investigate whether this relationship would also hold for T Cep, where our assigned effective temperatures were generally cooler than those of RY Cep and SU Cam, we computed continuum flux ratios as described above for our T Cep spectra. In Figure 10 we combine those new data with our published data for RY Cep and SU Cam and show that a fourth-order polynomial gives a reasonable fit to the data for all three stars. The mean spread in effective temperatures with respect to this polynomial fit is ± 38 K, consistent with our estimated uncertainty in effective temperature. It therefore appears that there is a consistent relationship between effective temperature and this continuum flux ratio for three oxygen-rich Miras spanning a temperature range from 2600 K to 4000 K. Whether this relationship holds more widely remains to be seen.

## 9. Summary

We observed one complete pulsation cycle of the oxygen-rich Mira star T Cep using concurrent spectroscopy and photometry. We used our photometry to calibrate spectra in absolute flux and measured how flux in four Balmer emission lines varied during the cycle. Rather than peaking once around the phase of maximum flux in the pulsation cycle, as we saw in other Mira stars, the Balmer emission lines in T Cep peak twice during each cycle and the phase at which these peaks occur is different for each emission line. We established the likely spectral sub-type of each spectrum by comparison with M-giant spectra on the MK system and used published data to estimate its corresponding effective temperature. We found a consistent relationship for three oxygen-rich Miras between assigned effective temperature and a ratio of fluxes measured at two high points on the spectral continuum.

## 10. Acknowledgements

The author thanks the anonymous referee for a helpful report which has improved the paper. He is also grateful to Lee Anne Willson for her valuable comments and advice. This research made use of the AAVSO Photometric All-Sky Survey (APASS) and the AAVSO Variable Star Index (VSX). The software developed for this project made extensive use of the Astropy package and the efforts of many contributors to this valuable community resource are gratefully acknowledged. We are also grateful to Tim Lester for his PlotSpectra software which saw everyday use in this project.

This research was made possible through the use of the AAVSO Photometric All-Sky Survey (APASS), funded by the Robert Martin Ayers Sciences Fund and NSF AST-1412587.

---

[1] Space Telescope Science Institute. 2021, CALSPEC (https://www.stsci.edu/hst/instrumentation/reference-data-for-calibration-and-tools/astronomical-catalogs/calspec).